# CUSTOMERS' PERCEPTION OF M-BANKING ADOPTION IN KINGDOM OF BAHRAIN : AN EMPIRICAL ASSESSMENT OF AN EXTENDED TAM MODEL

Ali AlSoufi[1] and Hayat Ali[2]

[1]Department of Information Systems, University of Bahrain, Bahrain
[2] Department of Information Systems, University of Bahrain, Bahrain

***ABSTRACT***

*Mobile applications have been rapidly changing the way business organizations deliver their services to their customers and how customers can interact with their service providers in order to satisfy their needs. The use of mobile applications increases rapidly, and has been used in many segments including banking segment. This research aims at extending the Technology Adoption Model (TAM) to incorporate the role of factors in influencing customer's perception towards M-banking adoption. Furthermore, the extended TAM model was evaluated empirically to measure its impact on M-banking adoption in  of Bahrain. The model was evaluated using a sample survey of 372 customers. The results reveal that the intention to adopt mobile banking is mainly affected by specific factors which are: Perceived Usefulness and Ease of Use. On the other hand, some factors such as perceived cost and perceived risk did not show any affect on the users' intention to use mobile banking. The result of this research is beneficial for banking service managers to consider the factors that can enforce the Mobile Banking services adoption and increase the take-up of their mobile services.*

## 1. INTRODUCTION

Mobile application or as the public around the world refer to it "mobile app" is a software application designed to run on mobile devices. They are available through various application platforms such as (Android, IOS, BlackBerry, Windows 8, etc).

The main purpose behind mobile applications development was to help retrieve information such as email and weather information. However, the rapid demand for more and various apps has led to more verity in mobile application categories including games, GPS, factory automation and dedicated online app-discovery services, e-Government services [41] and online banking. These mobile applications became very popular and mobile users were Addicted to their use and are employed in different sectors including banking sector.

Mobile application in of Bahrain has been increasing rapidly, and expected to increases in the coming years in banking sector. Most of the banks in the have started to launch mobile banking services. With intensive banks competition and the popularity of mobile device use, there is an urgent need to understand the factors that would entice customers to adopt Mobile banking. Thus, understanding the essentials of factors that determine user Mobile banking adoption can provide great management insight into developing effective strategies to remain competitive and hold market share.

Hence, in this research we aimed to propose an extended Technological Adoption Model for Mobile Banking (ETAMMB) to incorporate the role of factors influencing customer's perception towards M-banking adoption. In addition, the extended TAM was evaluated empirically to measure its impact on M-banking adoption in Bahrain.

The rest of the paper is organized as follows: first literature review about the mobile banking adoption is presented, second the research model and hypotheses are discussed, third the research instrument and sample is explained, then the results of testing the hypotheses are discussed and finally the conclusion with some recommendations are proposed.

## 2. LITERATURE REVIEW

With the rapid growth of mobile phones, the mobile services become a promising alternative for many sectors including banking sector. However, in comparison to the whole banking transactions, the market of mobile banking still remains very small [1, 2] especially that its usage is not reflecting on the adoption and usage of mobile banking [3].

Internet banking and mobile banking are both electronic banking [4]. However, they differ in the channels to be used in delivering the services to customers [5]. Thus, customers using Internet banking are using computers that are connected to Internet, while customers using mobile banking are using wireless devices to do transactions [6].

Literature reveals that research on electronic banking has focused on Internet banking, whereas research focusing on mobile banking receives little attention [7, 8, 9]. Table (1) presents a summary of empirical and theory-based empirical research in mobile banking adoption that was presented by [3]

Table 1: Empirical and theory-based empirical research in mobile banking adoption [3]

| Authors | Theories | Sampling | Main Findings |
|---|---|---|---|
| Suoranta and Mattila [2003] | Bass diffusion model and IDT | 1253 samples drawn from one major Finnish bank by the postal survey in Finland | Information sources (i.e., interpersonal word-of-mouth), age, and household income significantly influence mobile banking adoption |
| Laforet and Li [2005] | Attitude. Motivation, and behavior | 300 respondents randomly interviewed in the streets of six major cities in China | Awareness, confidential and security, past experience with computer and new technology are salient factors influencing mobile banking adoption |
| Luam and Lin [2005] | Extended TAM | 180 respondents surveyed at an e-commerce exposition and symposium in Taiwan | Perceived self-efficacy, financial costs, credibility, easy-of-use, and usefulness had remarked influence on intention to adopt mobile banking |
| Laukkanen [2007] | Mean-end theory | 20 qualitative in-depth interviews conducted with a large Scandinavian bank customers in Finland | Perceived benefits (i.e, location free and efficiency) are main factors encouraging people to adopt mobile banking |
| Amin et al. [2008] | TAM | 156 respondents obtained via convenience sampling in Malaysia | Perceived usefulness, easy-of-use, credibility, amount of information, and normative pressure significantly influence the adoption of mobile banking |
| Natarjan et al. [2010] | Analytical hierarchy process | 40 data obtained from a bank in India | Purpose, perceived risk, benefits, and requirements are main criteria to influence people to choose banking channels. |
| Koenig-Lewis et al. [2010] | TAM and IDT | 155 consumers aged 18-35 collected via online survey in Germany | perceived usefulness, compatibility, and risk are significant factors, while perceived costs, easy-of-use, credibility, and trust are not salient factors |
| Sripalawat et al. [2011] | TAM and TPB | 195 questionnaires collected via online survey in Thailand | Subjective norm is the most influential factor, the following is perceived usefulness and self-efficacy |
| Dasgupta et al. [2011] | TAM | 325 usable questionnaires gathered from MBA students in India | Perceived usefulness, easy-of-use, image, value, self-efficacy, and credibility significantly affect intentions toward mobile banking usage |

## 3. RESEARCH MODEL AND HYPOTHESES IN MOBILE BANKING ADOPTION

There are many studies in Mobile banking that applied research models and frameworks traditionally used within the IS literature [10]. The next sections present some of these models and based on that present this particular research extended model.

### 3.1. Technology Acceptance Model

First, the Theory of Reasoned action (TRA) was proposed by [11] to predict the human and explain human behaviour in various domains. Then, [12] based on TRA model proposed a Technology Acceptance Mode (TAM). The original TAM was presented in terms of two important determinants for systems use that are perceived ease of use (PEOU) and perceived usefulness (PU). In this model attitude toward using (ATU) directly predicts users' behavioural intention to use (BI) which determines actual system use (AU).

Many researchers suggested that TAM needed to include additional variables to provide a stronger model [13]. An extension, TAM2 was proposed by [14], which included social influence processes (subjective norm, voluntarism, and image) and cognitive instrumental processes (job relevance, output quality, result demonstrability, and PEOU), but it omitted ATU due to weak predictors of either BI or AU.

## 3.2. Extended TAM Model for M-Banking (ETAMMB)

Previous research was reviewed to ensure that comprehensive list of measures were included. The factors that are considered in this research that affect the adoption of Mobile Banking in of Bahrain were selected from TAM studies [12], the Extended TAM studies [15] and the effectiveness evaluation study by [16]. The research model has been tested using fifteen hypotheses as shown in Fig 1. These hypotheses are as follows:

**H1:** Customer services have a positive effect on the perceived usefulness of mobile.
**H2:** Quality of mobile have a positive effect on the perceived usefulness of mobile.
**H3:** Alternatives have a negative effect on the perceived usefulness of mobile.
**H4:** Efficient Transactions have a positive effect on the perceived usefulness of mobile banking.
**H5:** Efficient Transactions have a positive effect on the perceived ease of use of mobile banking.
**H6:** Compatibility has a positive effect on the perceived ease of use of mobile banking.
**H7:** Self-efficacy has a positive effect on the perceived ease of use of mobile banking.
**H8:** Perceived cost have a negative effect on behavioural intention to use mobile banking.
**H9:** Perceived risk have a negative effect on behavioural intention to use mobile banking.
**H10:** Perceived usefulness have a positive effect on the behavioural intention to use mobile banking.
**H11:** Perceived ease of use has a positive effect on the behavioural intention to use mobile banking.
**H12:** perceived cost have a positive effect on the perceived usefulness of mobile banking.
**H13:** perceived cost have a positive effect on the perceived ease of use of mobile banking.
**H14:** perceived risk have a positive effect on the perceived ease of use of mobile banking.
**H15:** perceived risk have a positive effect on the perceived usefulness of mobile banking.

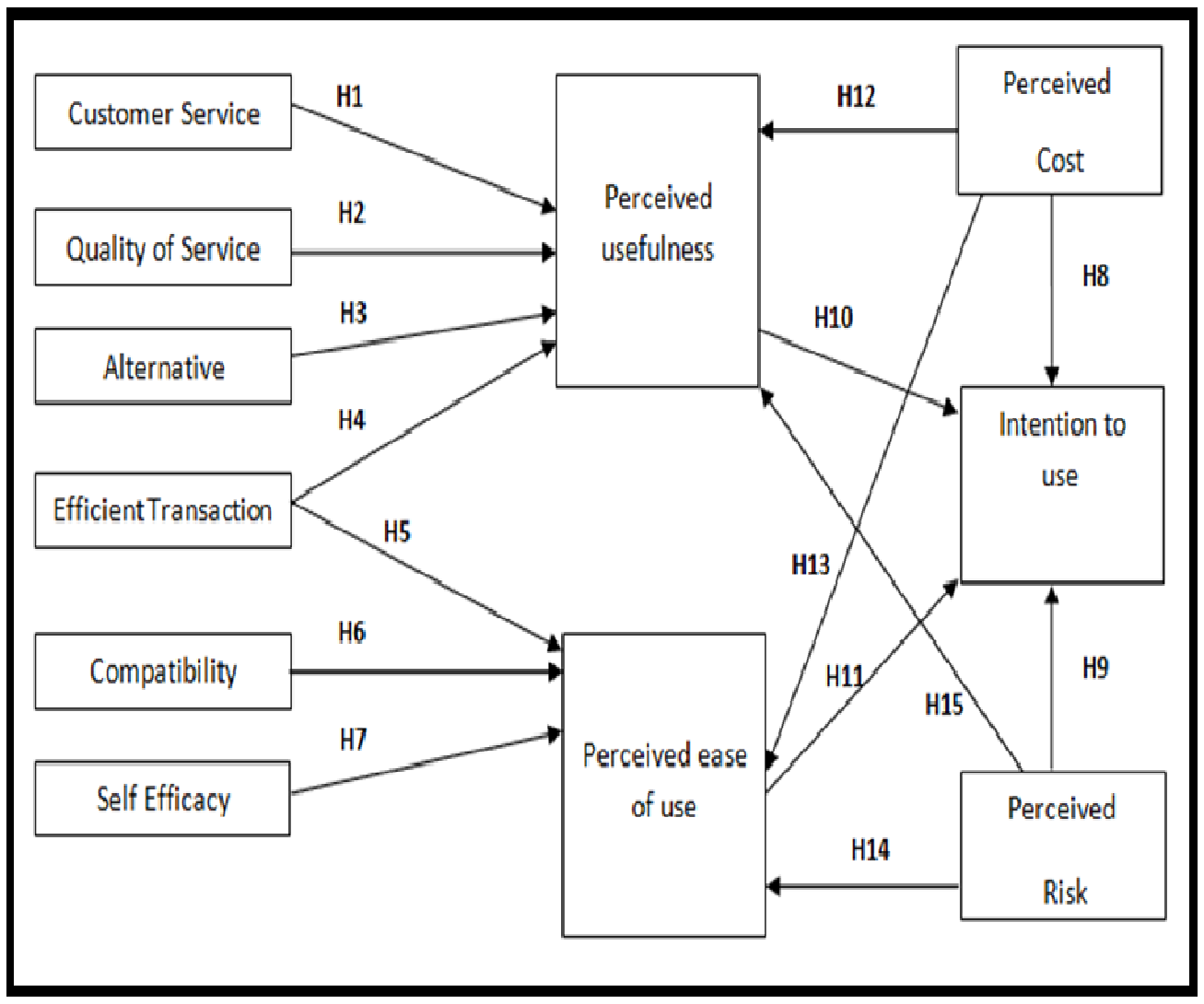

Figure1. Extended TAM Model for M-Banking (ETAMMB)

### 3.3. Definitions of Variables in the Model

**Customer service (SC):** Are characteristics used to evaluate the services offered by mobile banking to satisfy customer needs [17]. Customer service has four sub factors which are *Speed* [18] *Mobility Access* [19], *Advertising* and *Functions* [17].

**Quality of Service (QS):** Any service has a requirement and should achieve it. The measure of the quality depends on how the users or stakeholders perceived the services in terms of *Awareness* [20], Reliability [21], Accessibility*, Availability, and Accuracy, Responsiveness and Courtesy & helpfulness* [16]

**Alternatives (ALT):** This principle in regard to mobile banking will be used to efficiently express the environment of customer. And mobile banking could be one of possible way to do transaction. For example if the customer urgent to need to deal with the bank and no ATM machine or branch near to him, he can use mobile banking [17].

**Efficient Transaction (ET):** Any online services or transactions must be private and secure to insure the acceptance of user [16]. Five sub factors under this goal which are *Usability* [22], *Simplicity*, *Timeliness*, *privacy* [16], *Trust* [23], and *Security* [24].

**Self Efficacy (SE):** Self efficacy can be defined as "a judgment of one's ability to use a mobile banking service" [25] Self efficacy has three sub factors which are *Ability*, *Experience* [26] and *Knowledge* [27].

**Perceived Cost (PC):** The possible expenses of using Mobile banking many include equipments costs, access cost, and transaction fees [28].

**Perceived Risk (PR):** The risk regarded to the service itself [29]. It refers to the users' expectation of suffering a loss in the outcome of using Mobile Banking [30].

**Perceived Usefulness (PU):** Customer tends to use or not use a system to the extent they believe it will help them do their task well [12].

**Perceived Ease of Use (PEOU):** Perceived ease of use refers to the degree to which a person believes that engaging in online transactions via Mobile banking would be free of effort [12]. A system that is easy to use will accomplish tasks easily rather than system that is difficult to use [31].

**Compatibility(C):** That is the degree to which engaging in online transactions via Mobile banking is perceived as being consistent with the potential user's existing values, beliefs, previous experiences and current needs [32].

**Intention To Use (ITU):** Which refers to the user's likelihood to use online transactions through Mobile banking [33].

### 3.4. Research Instrument and Sample

In this research a questionnaire was employed to assess the developed model. Bahraini citizens who have experience, ability, or knowledge in using mobile applications, especially in using mobile banking were selected. Samples were chosen randomly to ensure that differences in responses to questions among different citizens. The researchers used Slovin's formula (n = N / (1 + Ne^2) [34] to obtain the random sample size that is 400 respondents in this research however 372 responses were collected. The survey questionnaire consisted of two parts. The first section was about the subject's demographic information. The second section was about subject's perception of each variable in the model. The second section asked each subject to indicate his or her degree of agreement with each item. Data were collected using a five point Likert-type scale.

### 3.5. Validity and Reliability of the Model Factors

In this research we attempted to examine the factors that affect consumers' adoption of Mobile banking by employing a modified TAM model. The TAM model is developed in order to verify the relations between the dependent variables and independent variables and test the hypotheses. SPSS analysis technique was used to assess the validity and reliability for each factor that affect the intention to use mobile banking services.

As [35] stated that the reliability is referring to the consistency of a measure, and a test is considered reliable if the tester get the same result repeated trails. There are many types of reliability including inter-ratter reliability, Test-retest reliability, parallel-forms Reliability and internal consistency reliability. In this research, the internal consistency reliability was tested as shown in Table I. It shows the results of alpha coefficients for each factor with reliability analysis. All factors are above 0.7 which indicate that good internal consistency of the questionnaire items, as [36] recommended that what is equal to 0.7 or above is acceptable.

Table1: Reliability of the model factors

| **Factors** | **Cronbach's Alpha** |
|---|---|
| Customer Services (CS) | 0.823 |
| Alternatives (ALT) | 0.701 |
| Self-Efficacy (SE) | 0.700 |
| Quality of Services (QS) | 0.828 |
| Efficient Transaction (ET) | 0.872 |
| Compatibility (COM) | 0.790 |
| Perceived Usefulness (PU) | 0.756 |
| Perceived Ease Of Use (PEOU) | 0.863 |
| Perceived Risk (PR) | 0.700 |
| Perceived Cost (PC) | 0.700 |
| Intention To Use (ITU) | 0.773 |

On the other hand, [35] defines the validity as the extent to which a test measures what it claims to measure. Using the factor aloading analysis, the values of component extracted were all above 0.5 which are at acceptable level.

### 3.6. Respondents' Profile

The number of female respondents in this sample was slightly higher than the number of males. Thus, 65.1% of the total respondents were females and 34.9% are males, this is because we distributed more questionnaire copies to females. Nearly half of the mobile banking users (50.8%) were aged between 18 and 24 years, so relatively the respondents were young especially as [37] asserted that young individuals are more likely to adopt internet banking while few mobile banking users were above 55 years old.

## 4. HYPOTHESES TESTING

In this research, hypothesis testing was performed on the basis of linear regression analyses. Linear regression is a method to find a relationship between one dependent variable and an independent variables [38].The independent variables and the dependent variable are integrated and tabulated in SPSS for hypothesis testing. Hypothesis testing is based and relies on the standardized coefficient significant. To support the hypothesis, the significant of the standardized coefficient should be below the 0.05 level. In order to calculate the r-path coefficient the following steps were followed:

1. *Independent variables:* customer service, Quality of Service, Efficient Transaction and alternatives are individually regressed against the dependent variable perceived usefulness (Hypotheses H1, H2, H3, and H4).
2. *Independent variables:* Efficient transaction, compatibility and self-efficacy are individually regressed against the dependent variable perceived ease of use (Hypotheses H5, and H6).
3. *Independent variables:* perceived cost, perceived risk, perceived usefulness and perceived ease of use, are individually regressed against the dependent variable intention to use (Hypotheses H7, H8, H9 and H10).

Accordingly, the following results obtained:

**Customer Service and Perceived Usefulness:** Hypothesis 1, that is Customer Service has the strongest relationship with perceived usefulness as standardization coefficient that equals to 0.389, which is greater than the accepted rate 0.1 and the significant is 0.000, which is lower than the accepted rate 0.05.

**Quality of Service and Perceived Usefulness:** Hypothesis 2, that is Quality of Service has a negative relationship with perceived usefulness that is rejected because of standardization coefficient that equals to -0.040 that is lower than the accepted rate 0.1 and the significant is 0.532 that is greater than 0.05. Our result contract with what [39] concluded with; in term of the rejection of this hypothesis, as he said that banks should be accessible, that is a sub factor under quality of service factor, in order to influence customer's perception of the usefulness.

**Alternative and perceived usefulness:** Hypothesis 3, that is alternative has a positive effect on perceived usefulness, is rejected because of standardization coefficient that equals to 0.070 which is lower than the accepted rate 0.1 and the significant is 0.195 that is higher than 0.05. Our result is consistent with the finding of [17] as almost all of the respondents in his sample are live in cities getting the latest technologies applied with banking service and prefer to go to the bank to do their transactions rather than using a mobile.

**Efficient Transactions and Perceived Usefulness:** Hypothesis 4, that is efficient transactions has a positive relationship with perceived usefulness that is accepted because of standardization coefficient that equals to 0.325 that is greater than the accepted rate 0.1 and the significant is 0.000 that is lower than 0.05. Our result is consistent with the finding of [39] as he emphasised that Mobile banking services should be secured and trustworthy, security and trust are sub factors of efficient transaction original factor, to influence customer's perception of the usefulness.
From the above four hypotheses (H1, H2, H3, H4): customer service (H1) has the strongest effect in the perceived usefulness, because standardization coefficient is equal 0.389 that is greater than the other factors that are affect perceived usefulness.

**Efficient Transaction and Perceived Ease of Use:** Hypothesis 5, that is transaction has a positive relationship with perceived ease of use that is accepted because of standardization coefficient that equals to 0.185 that is greater than the accepted rate 0.1 and the significant is 0.000 that is lower than 0.05.

**Compatibility and Perceived Ease of Use:** Hypothesis 6, that is compatibility has the strongest effect on perceived ease of use, is accepted because of standardization coefficient that equals to 0.460 that is greater than the accepted rate 0.1 and the significant is 0.000 that is lower than 0.05. This suggests that when the mobile banking services fit with customer life style, then a better ease of use the mobile banking will become.

**Self-Efficacy and Perceived Ease of Use:** Hypothesis 7, that is self-efficacy has a positive effect on perceived ease of use, is accepted due to standardization coefficient value that equals to 0.255 which is greater than the accepted rate 0.1 and the significant is 0.000 that is lower than 0.05. Our result is consistent with the finding of [17] in terms of the positive impact.

From the above three hypotheses (H5, H6, H7): Compatibility has a strongest effect on the perceived ease of use, because standardization coefficient that equals to 0.460 that is the greatest value compared with other variables that effect perceived ease of use.

**Perceived cost and Intention to Use:** Hypothesis 8, that is perceived cost has a negative relationship with intention to use , and this hypothesis is rejected because of standardization coefficient that equals to -0.059 which is lower than the accepted rate 0.1 and the significant is 0.136 that is greater than 0.05. The main reason for this result is that the respondents may not aware about the cost of mobile banking services because they did not use it as a banking option yet as Shi said that. Another reason might be that the banks are not providing their customers with enough information about mobile banking cost.

**Perceived Risk and Intention to Use:** Hypothesis 9, that is perceived risk has a positive relationship with intention to use, but is rejected because of standardization coefficient that equals to 0.091 which is lower than the accepted rate 0.1. One possible reason might be that users may not be aware about the risk of mobile banking, because not use it as banking option yet as [17] found.

**Perceived Usefulness and Intention to Use:** Hypothesis 10, that is perceived usefulness has a positive relationship with intention to use that is accepted because of standardization coefficient that equals to 0.287 which is greater than the accepted rate 0.1 and the significant is 0.000 that is lower than 0.05. Our result consistent with the result of [17] and his references studies in term of positively relationship. This suggests if customers believe that mobile banking is useful, faster and easier option to do their transaction rather than visit the bank, then they will use mobile banking.

**Perceived Ease of Use and Intention to Use:** Hypothesis 11, that is perceived ease of use has a strongest relationship with intention to use that is accepted because of standardization coefficient that equals to 0.442 which is greater than the accepted rate 0.1 and the significant is 0.000 that is lower than 0.05. Our result consistent with the result of [17] and his references studies in term of positively relationship. This suggests if customer believe that mobile banking is easy to use, and then will use mobile banking. On the other hand, this result is not consistent to the results obtained by [40] who found out that that Perceived Ease of Use affects Perceived Usefulness but does not impact on Attitude towards adoption.

From the above four hypotheses (H8, H9, H10, H11): Perceived ease of use had a strongest effect on the intention to use mobile banking; because it has the highest standardization coefficient equal 0.442 compared with perceived cost, perceived usefulness and perceived risk that effect intention to use.

**Perceived Cost and Perceived Usefulness:** Hypothesis 12, which is perceived cost has no relationship with perceived usefulness that is rejected because of standardization coefficient that equals to -0.109 which t is lower than the accepted rate 0.1.

**Perceived Cost and Perceived Ease of Use:** Hypothesis 13, that is perceived cost has no relationship with perceived ease of use that is rejected because of standardization coefficient that equals to -0.159 which is lower than the accepted rate 0.1 and the significant is 0.000 that is lower than 0.05.

**Perceived Risk and Perceived Usefulness:** Hypothesis 14, that is perceived risk has a strong relationship with perceived usefulness that is accepted because of standardization coefficient that equals to 0.403 which is greater than the accepted rate 0.1 and the significant is 0.000 that is lower than 0.05. Logically, when mobile banking is low risky then it will enable the people to use it more.

**Perceived risk and perceived ease of use:** Hypothesis 15, that is perceived risk has a strong relationship with perceived ease of use that is accepted because of standardization coefficient that equals to 0.432 which is greater than the accepted rate 0.1 and the significant is 0.000 that is lower than 0.05. Thus, if mobile banking is easy to use, it lowers the risk of making mistakes and transfer of wrong transactions. Table 2 summarizes the above results.

Table 2: Hypotheses testing results

| Hypothesis | Standardization coefficient | significant | Acceptance/ Rejection |
|---|---|---|---|
| **H1** | 0.389 | 0. 000 | Accepted |
| **H2** | -0.040 | 0.532 | Rejected |
| **H3** | 0.070 | 0.195 | Rejected |
| **H4** | 0.325 | 0.000 | Accepted |
| **H5** | 0.185 | 0.000 | Accepted |
| **H6** | 0.460 | 0.000 | Accepted |
| **H7** | 0.255 | 0.000 | Accepted |
| **H8** | -0.059 | 0.136 | Rejected |
| **H9** | 0.091 | 0.037 | Rejected |
| **H10** | 0.287 | 0.000 | Accepted |
| **H11** | 0.442 | 0.000 | Accepted |

To conclude all the relations, the study shows that the intention to use mobile banking is mainly affected by the perceived usefulness, perceived risk and ease of use. Figure 2 shows the research model describing the relations between each factor.

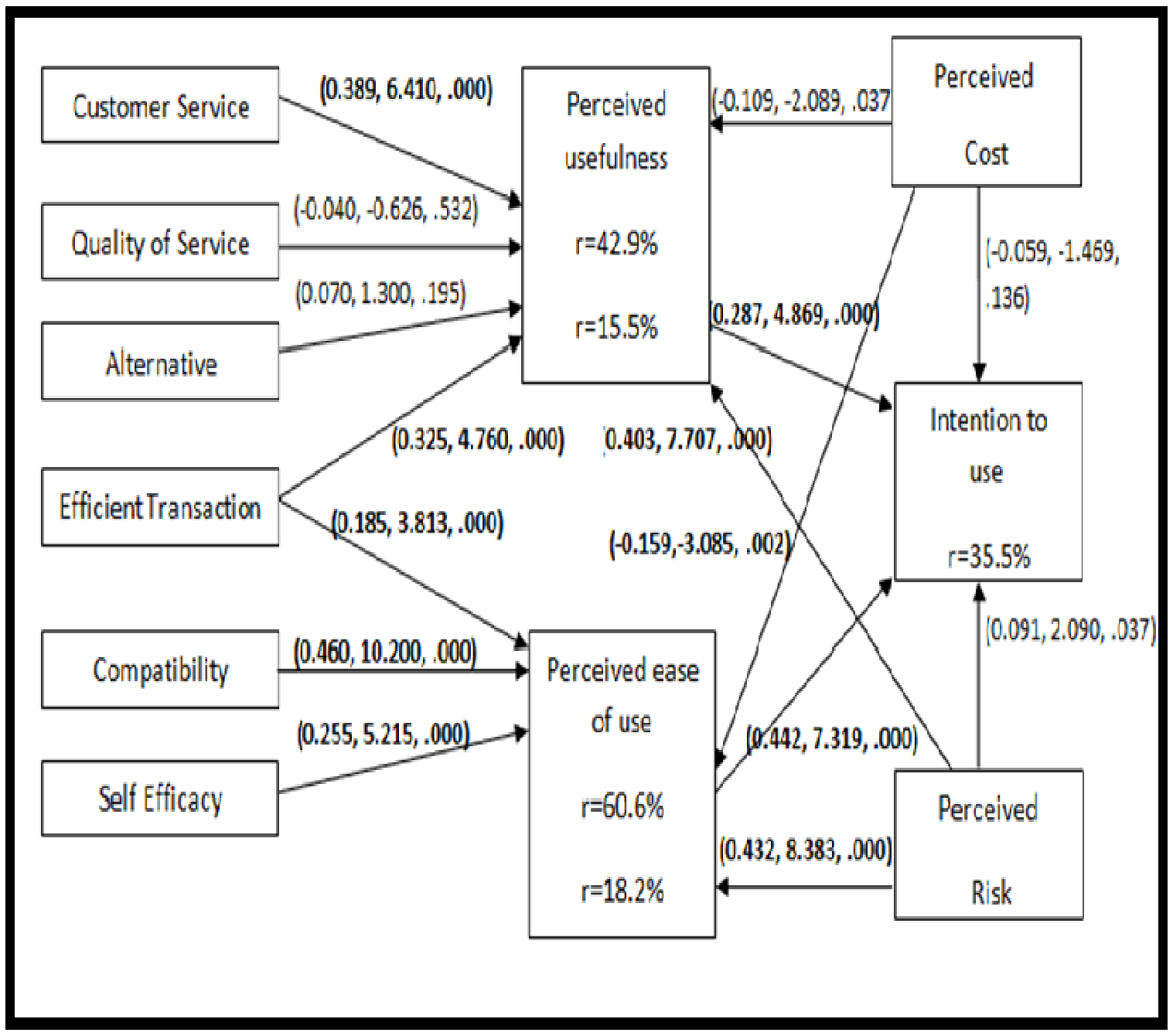

Figure 2. The research model and the relationship between the factors

## 5. CONCLUSION

This research presented an extended Technology Adoption Model (TAM) to incorporate the role of factors in influencing customer's perception towards M-banking adoption. In addition, the results of measuring this model empirically were presented to measure the factors' impact on M-banking adoption in of Bahrain

It was concluded after the analysis process, that the research's results are in line with the [17] study in New Zealand context. The research model studied four perceptions of users which are *perceived usefulness*, *perceived ease of use*, *perceived cost* and *perceived risk*. It was found that *perceived usefulness* strongly affected by *customer service*, with standardization coefficient of 0.389, and *efficient transaction* factor, with standardization coefficient that equals to 0.325. As well as, *perceived ease of use* affected strongly by *compatibility*, about 0.460 standardization coefficient, and *self-efficacy* with standardization coefficient of 0.255.

Moreover, *perceived usefulness* and *perceived ease of use* are affecting *intention to use* directly and strongly, in which *perceived ease of use* (standardization coefficient=0.442) has more impact than *perceived usefulness* (standardization coefficient=0.287). The other factors such as *perceived cost* (Standardization coefficient= -0.059) and *perceived risk* (Standardization coefficient= -0.091) have no effect on the intention to use mobile banking directly, but affecting indirectly through examined the relationship with *perceived usefulness* and *ease of use*. Only *perceived risk* has indirect relationship with intention to use through *perceived usefulness* (standardization coefficient=0.403) and *perceived ease of use* (Standardization coefficient=0.432).

Mobile penetration is in Bahrain now more than 17%. This makes a good opportunity to increase the level of adoption of Mobile banking, which is still in its early stages. We should take advantages of supported factors identified in this study and give attention to the unsupported factors. According to the research findings, certain factors are identified as the most critical while affect the intention to use for mobile banking in Bahrain. The following identified factors must be considered by banks in the of Bahrain to enhance their customer services and increase their customer base:

- Mobile banking usefulness has to be continuing improved in order to match the user interfaces of elderly, retailers and others leading the curve with mobile technology.

- Attention should be given to the risks which that could affect day-to-day transactions performed through mobile devices. Thus, it should be eliminated or reduced in order to enhance customers trust in the banking services being offered.

- To motivate customers to adopt this technology, the bank should try to increase the level of service expansion periodically as it should offer versatility in its offerings.

- Technical infrastructure of mobile banking services should be sophisticated and developed in order to ensure reliable and timely offering of services to customers.

- New functionalities a bank should bush it up to continue improve customers overall mobile experience and allow them to access most critical information.

As future works, this research could be expanded in terms of number of respondents to include different categories such as age, gender, nationality, etc. Furthermore the study could go for a better understanding on other segments of the industry such as M-Health, M-Education and M-Government to support mobile technology and its services and abilities. Open-ended questions may also be added to a future survey in order to provide deeper insight of customers' perception toward adoption of M-Banking and for a better generalization.

## ACKNOWLEDGMENT

This paper is one of the outcomes of the mobile apps research project that has been supported thankfully by the Deanship of research at University of Bahrain. Additionally we would like to thank Lulwa Ali, Hana Mohamed, Maryam Abdulla and Welaya Ali for their great efforts in data gathering through various interviews and literature search. They all had great role in bringing this work to the light.

## Authors

**Ali AlSoufi** is an assistant professor at University of Bahrain. He has earned his PhD in computer science in 1994 from Nottingham University, UK. He has worked for Bahrain Telecom Co for 8 years as a Senior Manager Application Programme where he overlooked number of mega IS Application projects. Dr Ali worked at Arab Open University as the head of IT program and Assistant Director for Business Development during 2007-2010, while working as a consultant for Bahrain e-Government Authority (EGA) in the area of Enterprise Architecture. He is also an active member of the Bahrain National ICT Governance Committee.

**Hayat Ali** is an assistant professor at University of Bahrain. She has earned her PhD in Information System in 2010 from the University of Manchester, UK. She has 11 years of experience in IS field. She has number of publications in various fields of Information Systems including the e-commerce, e-business, e-government and e-democracy.